\documentclass[letters,fleqn,usenatbib]{mnras}

\usepackage[british]{babel}
\usepackage{newtxtext}
\usepackage[slantedGreek]{newtxmath}

\usepackage[T1]{fontenc}

\usepackage{graphicx}	
\usepackage{amsmath}	
\usepackage{amssymb}	

\title[Stellar mass function of clumps at high redshift]{First constraints on the stellar mass function of star-forming clumps at the peak of cosmic star formation}

\author[M. Dessauges-Zavadsky and A. Adamo]{
Miroslava Dessauges-Zavadsky,$^{1}$\thanks{E-mail: miroslava.dessauges@unige.ch}
and Angela Adamo$^{2}$
\\
$^{1}$Observatoire de Gen\`eve, Universit\'e de Gen\`eve, 51 Ch. des Maillettes, CH-1290 Versoix, Switzerland\\
$^{2}$Department of Astronomy, Oskar Klein Centre, Stockholm University, AlbaNova University Centre, SE-106 91 Stockholm, Sweden
}

\date{Accepted June 15, 2018. Received YYY; in original form ZZZ}

\pubyear{2015}

\begin{document}
\label{firstpage}
\pagerange{\pageref{firstpage}--\pageref{lastpage}}
\maketitle
%
\begin{abstract}
Star-forming clumps dominate the rest-frame ultraviolet morphology of galaxies at the peak of cosmic star formation. 
If turbulence driven fragmentation is the mechanism responsible for their formation, we expect their stellar mass function to follow a power-law of slope close to $-2$. We test this hypothesis performing the first analysis of the stellar mass function of clumps
hosted in galaxies at $z\sim 1-3.5$. The clump sample is gathered from the literature with similar detection thresholds and stellar masses determined in a homogeneous way. To overcome the small number statistics per galaxy (each galaxy hosts up to a few tens of clumps only), we combine all high-redshift clumps. The resulting clump mass function follows a power-law of slope $\sim -1.7$ and flattens at masses below $2\times 10^7\,M_{\sun}$. By means of randomly sampled clump populations, drawn out of a power-law mass function of slope $-2$, we test the effect of combining small clump populations, detection limits of the surveys, and blending on the mass function. Our numerical exercise reproduces all the features observed in the real clump mass function confirming that it is consistent with a power-law of slope $\simeq -2$. This result supports the high-redshift clump formation through fragmentation in a similar fashion as in local galaxies, but under different gas conditions. 
\end{abstract}

\begin{keywords}
galaxies: evolution -- galaxies: high-redshift -- galaxies: structure
\end{keywords}



\section{Introduction}

High-redshift galaxies with clumpy morphologies seen in {\it Hubble} Space Telescope (HST) rest-frame ultraviolet (UV) images have first been reported back to \citet{cowie95}. 
Since then, \citet{guo15} have found that about 60\% of galaxies at the peak of the cosmic star formation ($z\sim 2$) are clumpy.
The formation process of the observed UV-bright clumps with stellar masses spread over $M_*^{\rm clump} \sim 10^{5.5}\,M_{\sun}$ to $10^{10.5}\,M_{\sun}$ and in most cases unresolved kiloparsec sizes \citep{forster11,guo12,guo18,elmegreen13,adamo13,wuyts14,soto17} is still debated between an ex-situ or in-situ origin. Recently, \citet{shibuya16} have shown that the fraction of clumpy galaxies is evolving with redshift, such that it rises from $z\sim 8$ to a peak around $z\sim 2$, and then declines to the present time. Their finding brings a strong argument against ex-situ clumps originating in interactions/mergers, since the clumpy galaxy fraction evolution is inconsistent with the observed and simulated evolutionary trends of both the major and minor mergers \citep{hopkins10,lotz11,rodriguez15}.

The in-situ clump formation trigged by the fragmentation of gaseous disks, subject to violent instabilities caused by intense cold gas accretion flows, currently is the most popular scenario to explain the origin of high-redshift star-forming clumps \citep{dekel09}. First, it is successfully reproduced by numerical simulations \citep[e.g.,][]{ceverino12,bournaud14,tamburello15,behrendt16,mandelker17}. Second, observational support in favor of rotation-dominated, highly turbulent, strongly star-forming, gas-rich, and marginally stable 
disks at high redshift is now well established \citep[e.g.,][]{forster09,tacconi13,wisnioski15,dessauges15,dessauges17b,harrison17,turner17,girard18}. 

Another evidence of the in-situ formation of high-redshift clumps via fragmentation resides in their stellar mass function. Indeed, stellar mass functions of young star clusters in nearby galaxies are commonly used to constrain their formation and how they are linked to the overall star formation process in galaxies. There is a growing consensus that star clusters form from a universal initial cluster mass function found to be a power-law distribution of the form $dN/dM \propto M^{-\alpha}$ with the index $\alpha\approx 2.0\pm 0.3$ \citep[e.g.,][]{gieles06,chandar14,adamo17}. This can be understood in the framework where turbulence is one of the driving mechanisms which governs the star formation by inducing the fragmentation via turbulent cascade. Because turbulence is a scale-free process, both gas and stars are expected to follow continuum density distributions that are described by log-normal functions \citep{elmegreen06,hopkins13,guszejnov18}. 

A meaningful exploitation of the stellar mass function of high-redshift clumps relies on the reliability of their derived stellar masses. In \citet{dessauges17a} we have pointed out how the lack of spatial resolution and sensitivity is affecting the determination of the intrinsic clump masses. The limiting resolution larger than 1\,kpc in high-redshift field/non-lensed galaxies, which is the best resolution achievable with HST at $z>1$,
yields blending effects and leads to a factor of $\lesssim 2-5$ increase in the clump masses \citep{tamburello17,cava18}. However, the sensitivity threshold used for the clump selection affects the inferred masses even more strongly than spatial resolution, biasing the detection of clumps at the low-mass end \citep{dessauges17a}. Using H$\alpha$ mock observations obtained from high-resolution hydrodynamical simulations of clumpy disk galaxies from \citet{tamburello15} post-processed with radiative transfer, we have been quantitatively able to evaluate that, indeed, the blending effect on the inferred clump masses is typically negligible in comparison to the sensitivity effect \citep{tamburello17}.


In this Letter, we use a well controlled sample of clumps hosted in high-redshift star-forming galaxies compiled from the literature (Section~\ref{sect:sample}) to derive their stellar mass function (Section~\ref{sect:results}). Its best-fit gives a power-law slope close to 2 in agreement with the scenario where star formation proceeds in a scale-free hierarchical fashion as a result of a turbulence-dominated interstellar medium (Section~\ref{sect:discussion}).

\begin{figure*}
\centering
\includegraphics[width=8.2cm,clip]{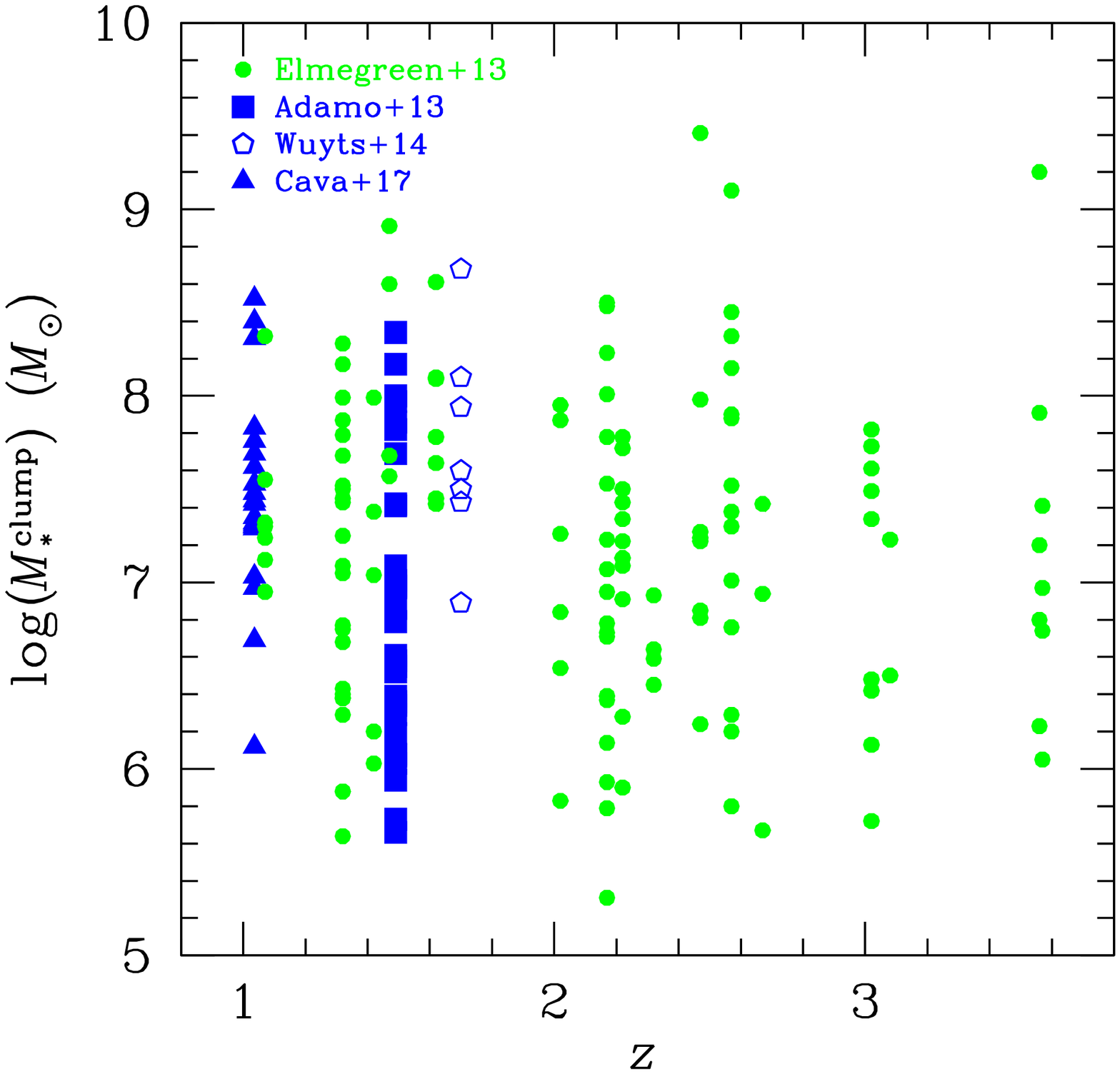}
\includegraphics[width=8.2cm,clip]{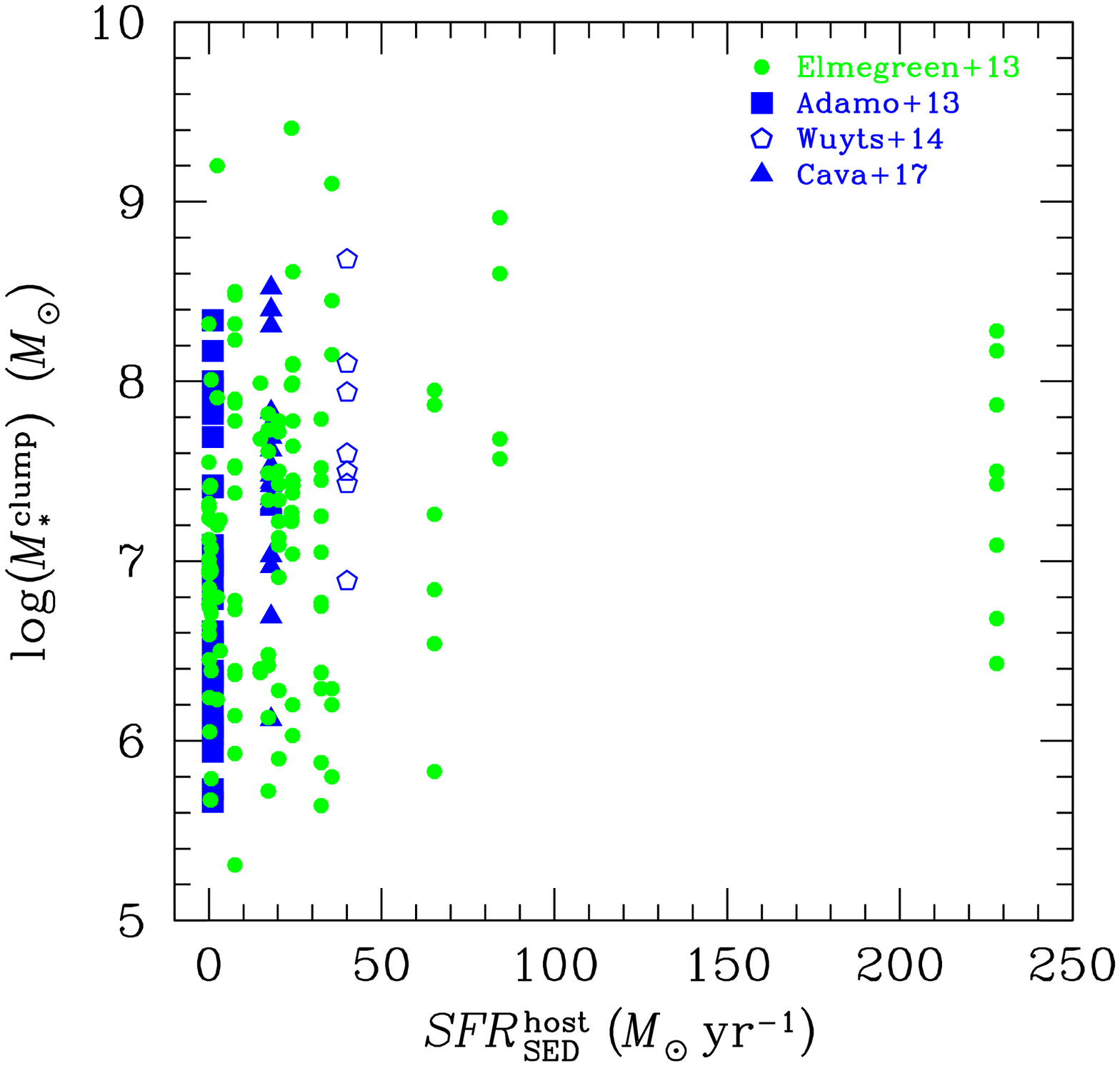}
\caption{Stellar masses of clumps plotted per redshift (left) and per star formation rate (right) of the host galaxy. The symbols correspond to different datasets: the blue squares, pentagons, and triangles refer to lensed galaxies, and the green dots to field/non-lensed galaxies.}
\label{fig:redshift-SFR}
\end{figure*}

\section{Description of the data sample}
\label{sect:sample}

In \citet{dessauges17a} we have compiled from the literature a sample of UV-bright stellar clumps hosted in distant star-forming galaxies at $1\lesssim z \lesssim 3.5$. Here we restrict this compilation to host galaxies from \citet{guo12}, \citet{elmegreen13}, \citet{adamo13}, and \citet{wuyts14} with available multi-band HST imaging that enables accurate clump stellar mass determinations from the spectral energy distribution (SED) modeling. We add to this compilation the clumps recently identified by \citet{cava18} in the Cosmic Snake, a strongly lensed star-forming galaxy at $z=1.036$. All the host galaxies were observed as part of the {\it Hubble} Ultra Deep Field \citep[HUDF;][]{beckwith06}, and the Cluster Lensing And Supernova survey with {\it Hubble} \citep[CLASH;][]{postman12} for two out of the three strongly lensed galaxies~--~with the exception of the lensed galaxy of \citet{wuyts14}. The five datasets reach very comparable $3\,\sigma$ sensitivity limits, in 0.35\arcsec\ diameter apertures, of 30.25 and 29.55\,AB\,mag (and only slightly better for the lensed host galaxies when accounting for the gravitational magnification) in, respectively, the HST F775W/F814W $i$-band and F850LP $z$-band (\citealt{dessauges17a}~--~Table~1; \citealt{cava18}). These two bands trace the young stellar rest-frame UV emission at $z\sim 1-3.5$, and have thus been predominantly used to select high-redshift clumps.

The comparison of the $i$- and $z$-band magnitude distributions of clumps shows very similar 16th percentile magnitudes of $\sim 29.7$\,AB\,mag in four datasets, except in the \citet{guo12} sub-sample where clumps are $2-2.5$ magnitudes brighter than the fainter clumps in the other four datasets (see \citealt{dessauges17a}~--~Table~1 and Figure~2; \citealt{cava18}). This directly affects the mass completeness of the \citeauthor{guo12} clump sub-sample that ends up to be complete down to stellar masses about 10 times more massive than in the other clump sub-samples. The incompleteness of the \citeauthor{guo12} clump sub-sample at the low-mass end results from their conservative clump selection ($m_{\rm F850LP}\leq 27.3$\,AB\,mag) well above the depth of the HUDF $z$-band image. Moreover, the host galaxies of \citeauthor{guo12} are biased toward the UV/optical luminous (and hence massive) end of the star-forming galaxy luminosity distribution, since chosen to have available spectroscopic observations. This is not the case of the other host galaxies that have all been serendipitously selected for their clumpy morphology in the HST $i$-band images. They thus turn out to be, on the contrary, biased toward the UV/optical fainter end, representative of the more numerous star-forming galaxies at $z\sim 1-3.5$ with stellar masses below the characteristic $M_*^{\star} \simeq 4\times 10^{10}\,M_{\sun}$ \citep[e.g.,][]{ilbert13}. Because of their incomplete clump selection and the different selection applied to the host galaxies, we exclude the \citet{guo12} clump sub-sample in the forthcoming analysis of the stellar mass function of clumps at high redshift.

For similar reasons, we choose to not include the recently published catalogue from \citet{guo18} of UV-bright clumps selected in star-forming galaxies at $0.5\leq z < 3$ from the shallower CANDELS/GOODS-S field. Even when restricting their host galaxies in redshift to $z\geq 1$ and in stellar mass to $M_*^{\rm host} < M_*^{\star}$ to match our compilation, 
this study suffers from a two magnitudes brighter clump selection than the fainter clumps from \citet{elmegreen13}, \citet{adamo13}, \citet{wuyts14}, and \citet{cava18}. Indeed, the 16th percentile magnitude of the \citet{guo18} clump distribution is equal to $\sim 27.5$\,AB\,mag in the HST F814W and F850LP filters.

We are thus left with the well defined sample of 194 clumps hosted in 25 galaxies~--~three of which are strongly lensed galaxies. All the clump stellar masses have been derived in a homogeneous way from the multi-band HST photometry \citep[see][]{dessauges17a,cava18}, using the updated version of the {\it Hyperz} photometric redshift and SED fitting code \citep{schaerer10}. The stellar tracks from \citet{bruzual03} at solar metallicity and the \citet{chabrier03} initial mass function (IMF) have been adopted. We have allowed for variable star formation histories, parameterized by exponentially declining models with timescales varying from 10\,Myr to infinity. No minimum age has been imposed, and nebular emission has been neglected. In Figure~\ref{fig:redshift-SFR} we show the derived clump stellar masses per redshift and star formation rate ($\mathit{SFR}$) of the host galaxy. Between 3 and 30 clumps are identified per galaxy. The combination of all the high-redshift clumps results in a stellar mass distribution that peaks at $\log(M_*^{\rm clump}) = 7.45\,M_{\sun}$ with a median mass of $7.24\,M_{\sun}$ in log.

The 59 clumps found in the three strongly lensed galaxies may serve as the control sample, since less affected by blending effect \citep{dessauges17a,cava18}. 
Their mass distribution nicely compares at its peak, median, and spread values to the mass distributions of the whole clump sample and even the sample of clumps hosted in field/non-lensed galaxies with limited kiloparsec resolution. This is consistent with the estimates made by \citet{tamburello17} using simulations and by \citet{cava18} using observations of multiple lensed images of the same galaxy. Both works find that blending, at least on scales between 1~kpc to $\lesssim 100$~pc, has only a weak effect on the derived clump stellar masses (a factor of $\lesssim 2-5$).

Finally, the expected dust attenuation of the star-forming host galaxies at $1\lesssim z\lesssim 3.5$ with stellar masses $M_*^{\rm host} \leq 2\times 10^{10}\,M_{\sun}$ is typically much lower than $A_V \lesssim 1$\,mag \citep{dominguez14}. Its impact on clump masses is much smaller than 0.4\,dex and in most cases is within the clump mass uncertainty, thus affecting very marginally the mass completeness of the clump sample.


\section{Analysis}
\label{sect:results}

The number of observed clumps in each high-redshift galaxy is too small (up to a few tens) to allow us any possible constraint on the stellar mass function of clumps at the peak of cosmic star formation. To overcome small sampling we test what type of constraint we get by combining together clumps detected in all galaxies of our sample. 

\begin{figure*}
\centering
\includegraphics[width=8.5cm]{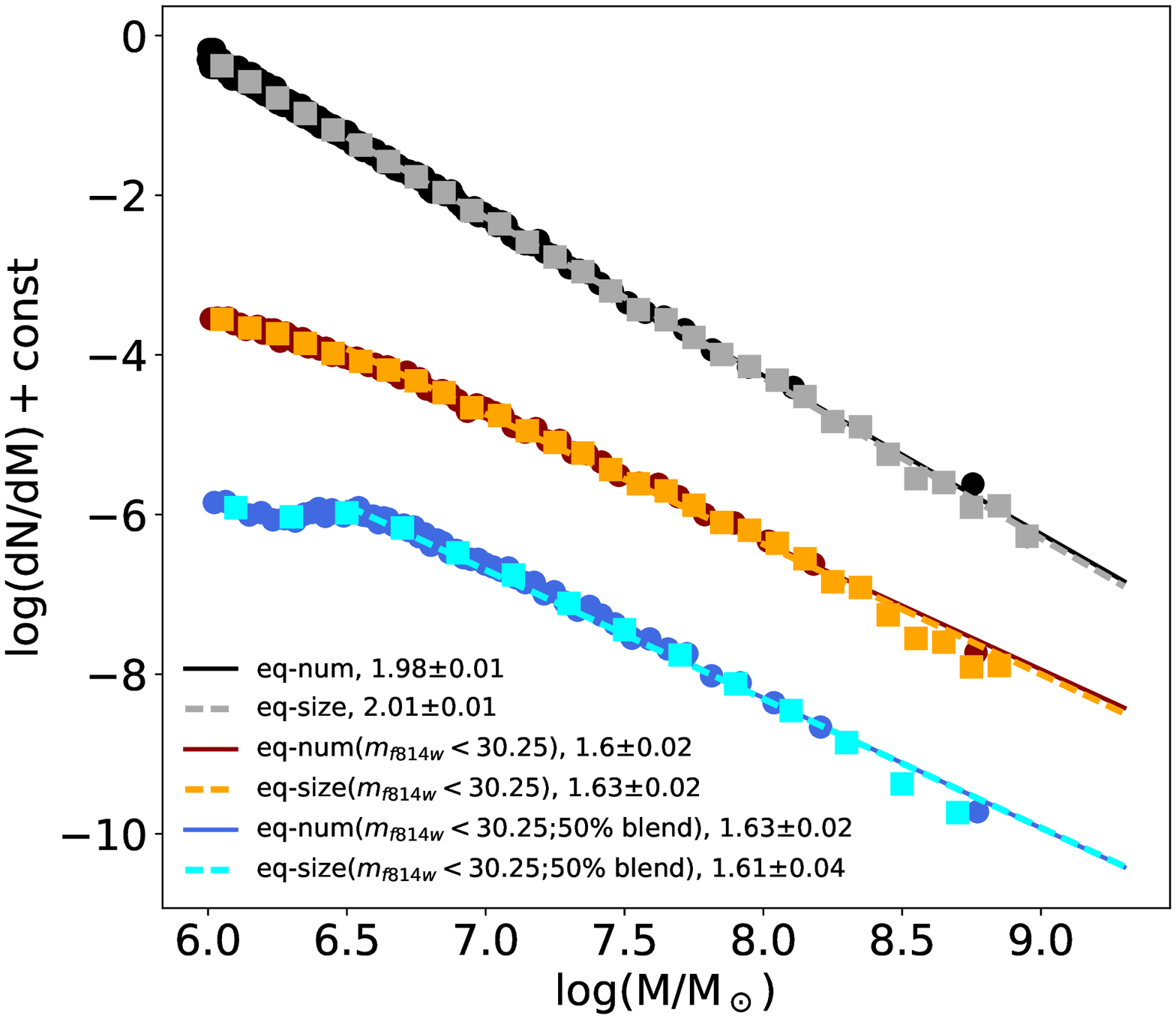}
\includegraphics[width=8.5cm]{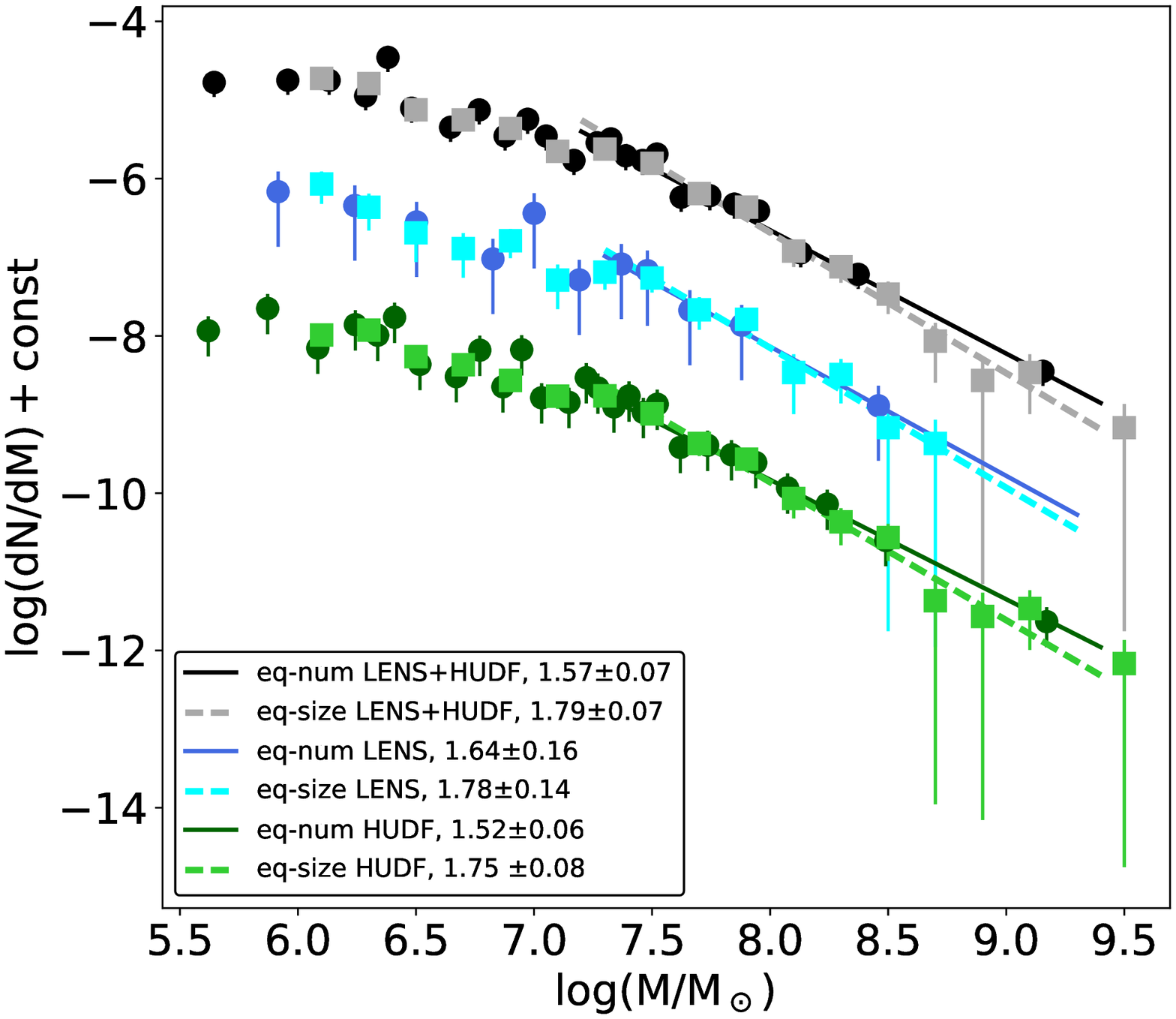}
\caption{Left: Clump mass distributions obtained from adding together populations of clumps with masses randomly drawn from a power-law distribution (see text). Right: Stellar mass distributions of clumps hosted in lensed galaxies \citep[LENS;][]{adamo13,wuyts14,cava18} and in field galaxies \citep[HUDF;][]{elmegreen13}. In both panels, a constant offset has been applied to the distributions to avoid their overlap. The dots and squares show the mass distributions using bins containing the same number of objects and fixed in size, respectively. The best fit to each distribution is illustrated with the solid and dashed lines, and the recovered slopes and uncertainties are listed in the inset. In the left panel, the fit to each distribution is performed down to $\log (M)\geq 6.5\,M_{\sun}$, the mass at which the break from a continuous distribution appears when the blending effect is accounted for (blue--cyan distributions). In our Monte Carlo exercise we test the effect of combining together clump populations (black--gray distributions), removing clumps with luminosities below the detection limit of the surveys ($m_{\rm F814W}< 30.25$\,AB\,mag; dark red--orange distributions), and removing clumps with luminosities below the survey limit after randomly blending 50\% of the clumps in each population (blue--cyan distributions). In the right panel, we show the distributions of our entire clump sample (LENS$+$HUDF, black--gray distributions), and separating clumps in lensed (blue--cyan distributions) with respect to those in non-lensed galaxies (dark--light green distributions). The fit is performed in all the distributions for bins with $\log (M)\geq 7.3\,M_{\sun}$, i.e. before the break occurs.}
\label{fig:clump_MF}
\end{figure*}

Using Monte Carlo simulations, we create clump populations sampling the redshift and $\mathit{SFR}$ parameter space occupied by our host galaxy sample shown in Figure~\ref{fig:redshift-SFR}. Our grid contains redshifts from 1 to 3.5, in steps of 0.5, and $\mathit{SFR}$ of 1, 3, 20, 40, and 100\,$M_{\sun}\,\rm yr^{-1}$. We consider as the total stellar mass formed in clumps the 20\% of the total stellar mass formed in the galaxy multiplying the $\mathit{SFR}$ by a time lapse of 300\,Myr. The two values are motivated by both observations \citep{guo12,adamo13,cava18} and simulations \citep{tamburello15,oklopcic17}. We randomly sample a power-law mass function, $dN/dM\propto M^{-\alpha}$ where $\alpha=2$, with clump masses between $10^6\,M_{\sun}$ and $10^9\,M_{\sun}$ until reaching the total stellar mass assumed to be forming in clumps. To each clump we randomly assign an age between 1\,Myr and 300\,Myr. Knowing the redshift of each galaxy and the age of each clump, we convert the masses into observed luminosities in the HST F814W broadband filter. We use Yggdrasil models \citep{zackrisson11} to estimates the observed magnitudes, assuming a Kroupa IMF \citep{kroupa01}, solar metallicity stellar libraries and ionized gas, a gas covering factor of 0.5, and a continuous $\mathit{SFR}$ for 10\,Myr. In the left panel of Figure~\ref{fig:clump_MF}, we show the resulting clump mass distribution obtained by combing together the clump populations of our simulated galaxy sample (top distributions in black dots and gray squares) using different binning techniques (equal number of objects and equal bin size, respectively). A single power-law in the form of $\log (dN/dM) = -\alpha\times M + const$ fitted to both distributions produces a slope similar to
the initial assumed value of $\alpha=2$. We conclude that combining a large sample of populations with small clump numbers yet drawn from a power-law mass function, results in a distribution consistent
to the initial power-law. We would like to stress that adopting a different clump age interval and star formation history (between instantaneous or continuous for a longer timescale) in our Monte Carlo simulations induces only a change in the number of clumps to be detected (the younger clumps being the brighter), and thus affects the normalisation of the recovered clump mass distribution but not the slope of the fitted power-law.

To test the effect of sensitivity as a function of redshift, we then apply an average detection limit ($m_{\rm F814W} < 30.25$\,AB\,mag) to the magnitudes in the F814W filter of our diverse clump populations. The resulting combined mass distributions are plotted with dark red dots (bins containing equal number of objects) and orange squares (bins of equal size) in the left panel of Figure~\ref{fig:clump_MF}. As a result of the intrinsic detection threshold in the data, the fit to the combined mass distribution returns a significantly smaller index $\alpha\simeq 1.6$.

As a final step, we also take into account the effect of blending. We randomly blend a fraction of clumps in each host galaxy, thus the resulting clump population includes blended and single sources. Before creating the combined mass distributions we reapply the detection limit cut. The resulting clump mass function is illustrated in the left panel of Figure~\ref{fig:clump_MF} with blue dots (bins with equal number of objects) and cyan squares (bins of equal size). We have tested 30\% and 50\% blending and show the latter case in the figure. Blending is drastically removing sources from the overall distribution and, in particular, in the lower mass bins, causing a break followed by a flat distribution. However, we notice that limiting the fit to the bins up to the break ($\log (M) = 6.5\,M_{\sun}$) produces the same slope as for the previous test
($\alpha\simeq 1.6$), hinting that the effect of sensitivity across our clump population sample is significantly more severe than the effect of blending, 
in agreement with real observations \citep{dessauges17a}.

In the right panel of Figure~\ref{fig:clump_MF} we show the clump mass distributions obtained by combining the observed clump populations from \citet{elmegreen13} (the field/non-lensed host galaxy sample, referred to as HUDF in the caption), and from \citet{adamo13}, \citet{wuyts14}, and \citet{cava18} (the lensed host galaxy sample, referred to as LENS in the caption). The LENS and HUDF mass range distributions significantly overlap. We observe a break in the distributions
as seen in the most realistic mock sample, thus we fit the distributions up to the break (namely for $\log (M_*^{\rm clump})\geq 7.3\,M_{\sun}$). We notice that the break occurs at a stellar mass that is almost 1\,dex higher compared to the mock sample. The difference in the position of the break in the observed and simulated mass distributions is likely produced by a more complex completeness function in the observed data than in the simulated ones and, in part, by the different SED modeling prescriptions (stellar evolutionary tracks, star formation history, and the treatment of the ionized gas) used to derive the observed clump masses and to perform the Monte Carlo simulations.
\citet{dessauges17a} report a difference of up to +0.56\,dex in their inferred clump masses with respect to the ones published by \citet{adamo13}, the latter using the same SED prescriptions
as the ones adopted to perform the Monte Carlo simulations in this work.
We may also notice that while we apply a simplistic detection cut, that is the same for all the clump populations, observationally the detection limits are slightly different (up to $0.2-0.5$\,mag), they depend also on the crowding of a region (clumps located within the bright background of their host galaxy) and, for the lensed sample, on the differential magnification. Nonetheless, overall we observe the same features in both mass distributions of observed and simulated clump populations, suggesting that while the absolute masses are depending on model assumptions, the effects of sensitivity and blending reflect in the resulting mass distributions in a similar manner. The two different binning methods (shown by the dot and square distributions) produce slopes which are within $2\,\sigma$ from each other. These differences simply arise because our clump sample is not very numerous and thus suffers from small number statistics as discussed in \citet{maiz05}. The fit to the two samples together (LENS$+$HUDF black and gray symbols and lines) or separated (blue--cyan and dark--light green distributions), produces slopes which are flatter than $\alpha=2$, but very close to the ones obtained in our Monte Carlo exercise. Indeed, running 700 realisations of clump populations and fitting the slope of each resulting clump distribution after applying the detection limit threshold and the 50\% blending, the peak of the distribution of the recovered slopes agrees within $1\,\sigma$ uncertainty with the slopes derived for the clump observations, and this for the two binning techniques considered. When excluding the most massive clumps with masses above $10^9~M_{\sun}$, which may potentially have an external origin of accreted satellite galaxies \citep{dessauges17a,mandelker17}, the change in the slopes for the observed clump samples (LENS$+$HUDF, LENS, and HUDF) is minimal and remains largely smaller than the error on the slope determination.

As a last test, we have repeated the Monte Carlo exercise assuming a flat distribution in the $\log(M/M_{\sun})$ space (i.e.\ $dN/dM\propto M^{-1}$). We observe that the simulated mass distributions have slopes incompatible with the observed ones. We thus conclude that the observed clump mass distribution at high redshift is consistent with an initial power-law function with $\alpha=2$.


\section{Discussion and Conclusions}
\label{sect:discussion}

It has been observed \citep{guo15,shibuya16} that the fraction of galaxies hosting UV-bright stellar clumps significantly increases around the peak of the cosmic star formation ($z\sim 2$). Numerical simulations suggest that clump formation results from disk fragmentation driven by gravitational instability \citep[e.g.,][]{dekel09,ceverino12,bournaud14,tamburello15}. High-redshift observations show us that many clumpy galaxies are rotation-dominated, highly turbulent, strongly star-forming, gas-rich, and marginally stable disks \citep[e.g.,][]{forster09,tacconi13,dessauges15,patricio18}. In local galaxies, gas fragmentation is driven by turbulence. It results in star-forming regions (star cluster complexes) that are hierarchically organized from parsec to kiloparsec scales and that follow a stellar mass distribution close to a power-law of slope $\alpha\simeq 2.0$ \citep[e.g.,][]{elmegreen06}. 

Until now, however, the stellar mass function of star-forming clumps at $z>1$ has remained unconstrained. The major challenges reside in compiling a well controlled sample of high-redshift clumps with a careful handling of the spatial resolution effect and, most importantly, the sensitivity effect by considering clumps selected under a similar detection threshold, and with stellar masses derived in a homogeneous fashion. Moreover, the determination of the clump mass function suffers from the small number of clumps hosted in each high-redshift galaxy, between a few up to tens of clumps only. In this work we have proposed to combine together the detected clump populations of several host galaxies across the peak of the cosmic star formation ($1\lesssim z\lesssim 3.5$) to observationally probe the stellar mass function of clumps at high-redshift. We test our methodology by means of simulated clump populations randomly sampled out a power-law function of slope $\alpha=2$, assuming redshifts and $\mathit{SFR}$ 
similar to the parameter space covered by our observational host galaxy sample. We find that combining together small numbers of clumps formed in each galaxy will result in a mass function with the same slope as the initial one.
According to our numerical exercise, the sensitivity of the datasets plays the largest effect, yielding a flattening of the resulting clump mass function. If fitted up to the break caused by the blending effect, the output slope is not largely affected and, although flatter than 2, is compatible with the latter value. Our observed clump mass function, fitted down to the break, results in a power-law function of slope $\alpha = 1.6-1.8$. We find this value to be consistent with the slope $\alpha\simeq 2$. 

Our finding supports the scenario where clumps form out of turbulence driven fragmentation even at high redshift. There may hence be a continuity in the way the largest coherent star-forming units form in galaxies from $z=0$ \citep[e.g.,][]{grasha17} up to $z\sim 3.5$ (the range we have been able to probe so far). However, locally these largest star-forming units have typical stellar masses rarely above $\sim 10^6 \,M_{\sun}$ \citep{adamo13}. The observed increase of the velocity dispersions \citep[e.g.,][]{wisnioski15} and molecular gas fractions \citep[e.g.,][]{dessauges17b} together with the decrease of the galaxy sizes \citep[e.g.,][]{shibuya15} with redshift result in galaxies across the peak of cosmic star formation characterized by a more turbulent and denser interstellar medium and with stronger pressure. These different/more extreme gas conditions allow to form clumps up to two orders of magnitude more massive and sample the stellar mass function at higher ranges.


\section*{Acknowledgments}
We thank the organisers of the Sexten meeting ``Across the scales of star formation'' held in 2017, and the visiting program at the Department of Astronomy of the Stockholm University for inspiring and providing the means for this work. We are thankful to Daniel Schaerer and Antonio Cava for helpful discussions, Matteo Messa for sharing with us some python routines used in this work, and the anonymous referee for helpful suggestions that have improved this letter. M.D.-Z. rewards the Swiss National Science Foundation in the context of the STARFORM Sinergia Project. A.A. acknowledges the support of the Swedish Research Council (Vetenskapsr\aa det) and the Swedish National Space Board (SNSB).



\bsp	
\label{lastpage}
\end{document}